# Fast MCMC sampling for Markov jump processes and continuous time Bayesian networks

**Vinayak Rao**
Gatsby Computational Neuroscience Unit
University College London
vrao@gatsby.ucl.ac.uk

**Yee Whye Teh**
Gatsby Computational Neuroscience Unit
University College London
ywteh@gatsby.ucl.ac.uk

## Abstract

Markov jump processes and continuous time Bayesian networks are important classes of continuous time dynamical systems. In this paper, we tackle the problem of inferring unobserved paths in these models by introducing a fast auxiliary variable Gibbs sampler. Our approach is based on the idea of *uniformization*, and sets up a Markov chain over paths by sampling a finite set of virtual jump times and then running a standard hidden Markov model forward filtering-backward sampling algorithm over states at the set of extant and virtual jump times. We demonstrate significant computational benefits over a state-of-the-art Gibbs sampler on a number of continuous time Bayesian networks.

## 1 Introduction

Many applications require modelling the time evolution of a dynamical system. A simple and popular approach is to discretize time and work with the resulting discrete time model. Such models have been well studied in the time series modelling literature [Rabiner, 1989, Murphy, 2002]. Often however, the evolution of the system is asynchronous with a number of different time scales. In this situation, the dependencies of the resulting time-discretized model can be sensitive to the chosen time scale. A more natural approach is to work directly with the continuous time dynamical system. One of the simplest continuous time dynamical system is the Markov jump process [Çinlar, 1975], which can be extended to incorporate structure into its dynamics via continuous time Bayesian networks [Nodelman et al., 2002]. Such models find applications in fields ranging from systems biology [Gillespie, 1977] to genetics [Fearnhead and Sherlock, 2006], network intrusion detection [Xu and Shelton, 2010] and human-computer interaction [Nodelman and Horvitz, 2003].

A challenge towards using these models is the problem of inference, which typically cannot be performed exactly. Various variational [Nodelman et al., 2002, Nodelman et al., 2005, Opper and Sanguinetti, 2007, Cohn et al., 2010] and sampling-based approximations [Fearnhead and Sherlock, 2006, Hobolth and Stone, 2009, El-Hay et al., 2008, Fan and Shelton, 2008] have been proposed in the literature, but do come with problems: usually they involve some form of time discretization, matrix exponentiation, matrix diagonalization, or root finding, and can be expensive for large problems.

In this work we describe a novel Markov chain Monte Carlo (MCMC) sampler for Markov jump processes and continuous time Bayesian networks that avoids the need for such expensive computations, is computationally very efficient, converges to the posterior with few iterations, and does not involve any form of approximations (i.e. our sampler will converge to the true posterior). Our method uses auxiliary variables which simplify the structure of Markov jump processes and allow the use of hidden Markov model forward filtering-backward sampling algorithms to efficiently resample the whole state sequence.

In Section 2 we briefly review Markov jump processes. In Section 3 we introduce the idea of uniformization and describe our algorithm for the simple case of homogeneous Markov jump processes. In Section 4 we briefly review continuous-time Bayesian networks and extend our algorithm to that setting. In Section 5 we report experiments comparing our algorithm to a state-of-the-art Gibbs sampler for continuous-time Bayesian networks and to variational inference for Markov jump processes. We end with a discussion in Section 6.

## 2 Markov Jump Processes (MJPs)

A Markov jump process ($\mathbf{S}(t)$, $t \in \mathbb{R}_+$) (see for example [Çinlar, 1975]) is a stochastic process with right-continuous, piecewise-constant paths. The paths themselves take values in some countable space $\mathcal{S}$ which, as in typical applications, we assume is finite (say $\mathcal{S} = \{1, 2, ...n\}$). We also assume the process is homogenous, implying (together with the Markov property) that for all times $t', t \in \mathbb{R}_+$ and states $i, j \in \mathcal{S}$,

$$p(\mathbf{S}(t'+t) = j | \mathbf{S}(t') = i, \mathbf{S}(u); u < t') = [P_t]_{ji} \quad (1)$$

for some stochastic matrix $P_t$ which depends only on $t$. The family of transition matrices $(P_t, t \geq 0)$ is defined by a matrix $A \in \mathbb{R}^{n \times n}$ called the *rate matrix* or *generator* of the MJP. $A$ is the time-derivative of $P_t$ at $t = 0$, with

$$P_t = \exp(At) \quad (2)$$
$$p(\mathbf{S}(t'+dt) = j | \mathbf{S}(t') = i) = A_{ji} dt \quad (\text{for } i \neq j) \quad (3)$$

where (2) is a matrix exponential and $dt$ is an infinitesimal quantity. The off-diagonal elements of $A$ are non-negative, and represent the rates of transiting from one state to another. Its diagonal entries are $A_i \equiv A_{ii} = -\sum_{j \neq i} A_{ji}$ for each $i$ so that its columns sum to 0, with $-A_i$ characterising the total rate of leaving state $i$.

Consider a time interval $[t_{start}, t_{end}]$ and let $\pi$ be an initial distribution over states at time $t_{start}$. Then an MJP can be described by the following generative process over paths on this interval:

1. At the initial time $t_0 = t_{start}$, assign the MJP a state $s_0$ with $p(\mathbf{S}(t_0) = s_0) = \pi_{s_0}$. Let $i = 1$.

2. Draw $z \sim \text{Exp}(A_{\mathbf{S}(t_{i-1})})$ and let $t_i = t_{i-1} + z$.

3. If $t_i < t_{end}$, the MJP jumps to a new state $s_i$ at time $t_i$, with $p(\mathbf{S}(t_i) = s_i | \mathbf{S}(t_i-) = s_{i-1}) \propto A_{s_i s_{i-1}}$ for each $s_i \neq s_{i-1}$. Increment $i$ and go to step 2.

4. Otherwise, set $t_i = t_{end}$ and stop.

Let $N$ be the number of jumps in the MJP. Then the sequence of times $T = \{t_0, \ldots, T_{N+1}\}$ along with the sequence of states $S = \{s_0, \ldots, s_N\}$ completely characterize the MJP path $\mathbf{S}(\cdot)$. Their probability[1] under the generative process above is given by

$$p(S,T) = \pi_{s_0} \left( \prod_{i=1}^{N} |A_{s_{i-1}}| e^{A_{s_{i-1}}(t_i - t_{i-1})} \frac{A_{s_i s_{i-1}}}{|A_{s_{i-1}}|} \right) \\ \cdot e^{A_{s_N}(t_{N+1} - t_N)} \quad (4)$$

$$= \pi_{s_0} \left( \prod_{i=1}^{N} A_{s_i s_{i-1}} \right) \exp\left( \int_{t_{start}}^{t_{end}} A_{\mathbf{S}(t)} dt \right) \quad (5)$$

Note that the last term in (4) is the probability that the process remains in state $s_N$ until at least time $t_{N+1}$.

In this paper, we are concerned with the problem of sampling MJP paths over the time $[t_{start}, t_{end}]$ given observations of the state of the MJP at a discrete set of times.

---
[1]Technically this is a density. For simplicity, we do not make this explicit in this paper.

In the simplest case, we observe the state of the process at the boundaries $t_{start}$ and $t_{end}$. More generally, we are given the initial distribution over states $\pi$ as well as a set of $O$ noisy observations $X = \{X_{t_1^o}, \ldots X_{t_O^o}\}$ with likelihoods $p(X_{t_i^o} | \mathbf{S}(t_i^o))$ and we wish to sample from the posterior $p(\mathbf{S}|X)$, or equivalently $p(S, T | X)$.

A simple approach when the states at the boundaries are observed is rejection sampling: sample paths from the prior given the observed start state and reject those that do not end in the observed end state [Nielsen, 2002]. This can be extended to the case of noisy observations by importance sampling or particle filtering [Fan and Shelton, 2008]. However, a large state-space, a long time interval or an unlikely end state can result in large numbers of rejections or small effective sample sizes.

A second approach marginalizes over the infinitely many paths of the MJP in between observations using matrix exponentiation (2), and uses forward-backward dynamic programming to sum over the states at the finitely many observation times (see [Hobolth and Stone, 2009] for a review). Unfortunately, matrix exponentiation is an expensive operation that scales as $O(n^3)$, $n$ being the number of states. Moreover, the matrix resulting from matrix exponentiation is dense and any structure, e.g. sparsity, in the rate matrix $A$ cannot be exploited.

In this paper we describe an MCMC algorithm to sample from the posterior distribution. Our method scales as $O(n^2)$, does not require matrix exponentiation, and can easily exploit structure in the rate matrix. Moreover, we demonstrate that our sampler mixes very rapidly.

## 3 MCMC inference via uniformization

We first introduce the idea of *uniformization* [Jensen, 1953, Çinlar, 1975, Hobolth and Stone, 2009], which forms the basis of our sampling algorithm. For an MJP with rate-matrix $A$, choose some $\Omega \geq \max_j (-A_j)$. Let $W$ be an ordered set of times on the interval $[t_{start}, t_{end}]$ drawn from a homogenous Poisson process with intensity $\Omega$. Because the Poisson rate $\Omega$ dominates the leaving-rates of all states of the MJP, $W$ will, on average, contain more events than the jump times $T$ of the MJP. We can 'thin' the set $W$ by rejecting a number of events from $W$. In particular, letting $I$ be the identity matrix, observe that $B = I + \frac{1}{\Omega}A$ is a stochastic matrix; run a discrete time Markov chain with initial distribution $\pi$ and transition matrix $B$ on the times in $W$. This is a *Markov chain subordinated to the Poisson process*. It will assign a set of states $V$ to the times $W$. Unlike $S$, the set $V$ can have *virtual* jumps where a state jumps back to itself. Just as $(S, T)$ characterize a MJP path, $(V, W)$ also characterize a sample path of some piecewise-constant and right-continuous stochastic process on $[t_{start}, t_{end}]$. As the parameter $\Omega$ increases, the number of times in $W$ increases; at the same time the diagonal en-tries of $B$ increases, so that the number of self-transitions also increases. The following proposition shows that these two effects exactly compensate each other, so that the process characterized by $(V, W)$ is precisely the desired MJP:

**Proposition 1.** *For any $\Omega \geq \max_i (-A_i)$, $(S,T)$ and $(V,W)$ define the same Markov Jump Process $\mathbf{S}(t)$.*

*Proof.* Simply write down and compare the two probabilities, see [Hobolth and Stone, 2009, Çinlar, 1975] □

### 3.1 The MCMC algorithm

We adapt the uniformization scheme described above to construct an auxiliary variable Gibbs sampler. The only difference between $(S,T)$ and $(V,W)$ is the existence of an auxiliary set $U$ of virtual jumps in $(V, W)$. We proceed by first drawing this set of virtual jumps given $(S,T)$, as a result recovering the uniformized characterization $(V, W)$. Given $V$, the distribution of $W$ is simply a Markov chain so we can now perform simple HMM forward filtering-backward sampling, incorporating evidence from observations, to obtain a new state sequence $\tilde{V}$. Finally, dropping the virtual jumps in $(\tilde{V}, W)$ gives a new MJP path $(\tilde{S}, \tilde{T})$.

Consider sampling $U$ from an *inhomogenous* Poisson process with intensity $R(t) = \Omega + A_{\mathbf{S}(t)}$. This intensity is piecewise-constant, taking the value $r_i = \Omega + A_{s_i}$ on the interval $[t_i, t_{i+1})$. Define $u_i$ as the number of auxiliary times over this interval. The probability of $U$ is then

$$p(U|S,T) = \prod_{i=0}^{N} \frac{r_i^{u_i}(t_{i+1}-t_i)^{u_i} e^{-r_i(t_{i+1}-t_i)}}{u_i!} \cdot \frac{u_i!}{(t_{i+1}-t_i)^{u_i}}$$

$$= \left(\prod_{i=0}^{N} (\Omega + A_{s_i})^{u_i}\right) \exp\left(-\int_{t_{start}}^{t_{end}} (\Omega + A_{\mathbf{S}(t)}) dt\right) \quad (6)$$

The $u_i!$ in the numerator of the first equation arises because $U$ is an ordered set.

**Proposition 2.** *For any $\Omega \geq \max_i(-A_i)$, the Markov jump process $(S,T)$ with auxillary times $U$ is equivalent to the times $V$ sampled from the subordinating Poisson process along with the states $W$ assigned via the subordinated Markov chain. In other words, $p(S,T,U) = p(V,W)$.*

*Proof.* Multiplying (5) with (6), we see that $p(S,T,U) =$

$$\frac{\Omega^{|U|+N}}{e^{\Omega(t_{end}-t_{start})}} \cdot \pi_{s_0} \prod_{i=0}^{N}\left(1 + \frac{A_{s_i}}{\Omega}\right)^{u_i} \prod_{i=1}^{N} \frac{A_{s_i s_{i-1}}}{\Omega} \quad (7)$$

The first term is the probability of an ordered set of times under a homogenous Poisson process with rate $\Omega$, while the second is the probability of a sequence of states under a Markov chain with initial distribution $\pi$ and transition matrix $B = (I + \frac{1}{\Omega} A)$. These are just $W$ and $V$. □

Now we can incorporate the likelihoods of observations $X$ into the subordinated Markov chain $V$. In the interval $[w_i, w_{i+1})$, the MJP is in state $v_i$, so that the observations in this interval gives a likelihood term:

$$L_i(v_i) = \prod_{j: t_j^o \in [w_i, w_{i+1})} p(X_{t_j^o} | \mathbf{S}(t_j^o) = v_i) \quad (8)$$

Conditioned on the times $W$, $V$ is a Markov chain with likelihoods given above, so we can efficiently resample $V$ using the standard forward filtering-backward sampling algorithm. This cost of this is $O(n^2 |V|)$, quadratic in the number of states and linear in the length of the chain. Further any structure in $A$ (e.g. sparsity) is inherited by $B$ and can be exploited easily.

Let $\tilde{V}$ be the new state sequence. Then $(\tilde{V}, W)$ will correspond to a new MJP path $\tilde{\mathbf{S}}(t)$, characterized by $(\tilde{S}, \tilde{T})$ by discarding virtual jumps from $(\tilde{V}, W)$.

**Proposition 3.** *The auxillary variable Gibbs sampler described above has the posterior distribution $p(\mathbf{S}(t)|X)$ as its stationary distribution. Moreover, if $\Omega > \max_i(-A_i)$, the resulting Markov chain is ergodic.*

*Proof.* The first statement follows from the fact that the algorithm simply introduces auxiliary variables $U$ followed by conditional sampling of $V$ given $X$ and $W$. To show ergodicity, note that if $\Omega > \max_i(-A_i)$, then the intensity of the subordinating Poisson process is strictly positive. Consequently, there is positive probability (density) of sampling appropriate auxiliary jump times $U$ and to move from any MJP path to any other. □

Note that it is essential for $\Omega > \max_i(-A_i)$. For example, if all diagonal elements of A are equal to $\Omega$, then the subordinating Poisson process with have intensity 0, and consequently the set of jump times $T$ will never be changed by the sampler above. In fact the only dependence between successive samples of the Gibbs sampler is through the jump times, since the state sequence $\tilde{V}$ is independent of $V$ given $W$. By increasing $\Omega$, more auxiliary virtual jumps are introduced allowing the sampler to move to different jump times quickly, leading to faster mixing. Of course, the HMM chain is longer, leading to a linear increase in the computational cost per Gibbs iteration. Thus the parameter $\Omega$ allows a trade-off between mixing rate and computational cost. In all our experiments, we set $\Omega = \max_i(-2A_i)$; we find this works quite well, with the samplers typically converging after less than 5 iterations.

To end, we point out that the idea of sampling using uniformization is not new (see [Hobolth and Stone, 2009]). Existing methods produce *independent* samples of the number of subordinating Poisson events given observations. Consequently, these are computationally more expensive, requiring cubic time matrix operations and cannot take advantage of structure in $A$. In comparison, our method is MCMC based, with the number of Poisson events dependent across samples, but is significantly faster. More importantly, for complicated likelihood functions, sampling from this posterior distribution is often hard (see for example [Fearnhead and Sherlock, 2006]). Our auxiliary variable MCMC sampler simplifies this difficult posterior sampling, with the likelihood only entering through the emission matrix in the forward-backward algorithm. In the next section, we describe a novel Gibbs sampler for *continuous time Bayesian networks*, where the likelihoods are complicated by the 'observations' being over continuous time rather than discrete.

# 4 Continuous-time Bayesian Networks (CTBNs)

Continuous-time Bayesian networks (CTBNs), introduced in [Nodelman et al., 2002], are compact, multi-component representations of MJPs with structured rate matrices. Just as the familiar Bayesian network represents a probability table whose size is exponential in the number of variables with a series of smaller conditional probability tables, a CTBN represents a structured rate matrix with smaller conditional rate matrices. An $m$-component CTBN represents the state of an MJP with the states of $m$ nodes, $\mathbf{S}^1(t), \ldots, \mathbf{S}^m(t)$ in a directed (and possibly cyclic) graph $\mathcal{G}$. Each node acts as an MJP with a particular rate matrix which depends on the instantaneous states of its parents but not its children (we discuss the children's effect later). Since the parents are themselves MJPs with piecewise constant paths, the rate matrix is inhomogeneous and piecewise constant. The graph $\mathcal{G}$ and the set of rate matrices (one for each node and for each configuration of its parents) characterize the dynamics of the CTBN. Completing the specification of the CTBN is an initial distribution $\pi$ over the nodes, specified via some Bayesian network $\mathcal{B}$.

To sample from a CTBN over an interval $[t_{start}, t_{end}]$, one follows a generative process similar to that for MJPs:

1. At the initial time $t = t_{start}$, assign the entire CTBN a configuration $\mathbf{S}(t) \equiv (s_0^1, s_0^2, ...) \sim \pi$.

2. For each node $k$, let $A^t$ be its rate matrix at time $t$. Note that we have suppressed the dependence of $A^t$ on $k$ and the configuration of its parents. Draw $z^k \sim \text{Exp}(A^t_{\mathbf{S}^k(t)})$.

3. Let $K = \text{argmin}_k z^k$ be the first node to jump (unique with probability 1), and $t = t + z^K$ the jump time.

4. If $t < t_{end}$, node $K$ jumps to state $s$ at time $t$, with $p(\mathbf{S}^K(t) = s|\mathbf{S}^K(t-) = s') \propto A^t_{s,s'}$ for each $s \neq s'$. Other nodes stay in their current states. Go to step 2.

5. Otherwise, we reached time $t_{end}$ and stop.

## 4.1 Inference in CTBNs

While a CTBN can be interpreted as a simple MJP over an expanded state space, this state space is exponentially large, so that sampling algorithms (even our algorithm in Section 2) cannot be directly applied. To develop a tractable MCMC sampler, we consider Gibbs sampling where parts of the CTBN are kept fixed while others are sampled from their conditional distribution.

To do so, a careful interpretation of the conditional independences in CTBNs is in order. Note in particular that for any time $t$ and node $k$, given the configuration of its Markov blanket $\mathcal{MB}$ at time $t$, its state $\mathbf{S}^k(t)$ is *not* independent of the state of some other node $k'$. This is because the temporal dynamics of the network cause all nodes in the graph to become entangled [Nodelman et al., 2002]. That is, the present state of $k'$ tells us something about its previous states, which in turn tells us something about previous configurations of the Markov blanket $\mathcal{MB}$ and of node $k$, thus resulting in dependence with the current state of $k$.

However, given the entire sample paths of all nodes in the Markov blanket, node $k$ is independent of all other nodes in the network. This suggests a Gibbs sampling scheme where the sample path of each node is resampled given those of its Markov blanket. This was the approach followed by [El-Hay et al., 2008]. A complication arises, however, because even over an interval of time where the configuration of the node's Markov blanket remains constant, the conditional distribution of the node's sample path is not a piecewise homogeneous MJP. This is because of the sample paths of the node's children, which effectively act as observations that are continuously observed. Consequently, the MJP sampling algorithm of Section 2, which require observations at a discrete set of times, cannot be applied.

[El-Hay et al., 2008] described a matrix exponentiation based Gibbs sampler that repeatedly samples the time of the next transition and assigns the node a new state. In addition to calculating the matrix exponentials of many different matrices, their method has to discretize time: to obtain the time of the next jump, they perform a binary search on the time interval up to a specified accuracy. While they argue that this allows the user to specify a desired 'precision', it would be better to not resort to time discretization. We next adapt our uniformization-based algorithm to exactly sample paths from the inhomogeneous MJP given the conditional distribution of a node given its Markov blanket. Besides being exact, we demonstrate in our experiments that this can result in significant computational gains.

## 4.2 Auxiliary Variable Gibbs sampling for CTBNs

In this section, we describe a Gibbs sampling algorithm to simulate the CTBN posterior over an interval $[t_{start}, t_{end}]$, given a set of observations $X$ at times $\{t_1^o, ...t_O^o\}$. An iteration of the overall algorithm proceeds by performing

Gibbs updates on all nodes in the CTBN. In the following we describe the update step for a single node $k$. To avoid notational clutter, we suppress all references to the node index $k$, as well as dependence of rate matrices on the configurations of parents. Thus, we are given the complete sample paths of all nodes in node $k$'s Markov blanket $\mathcal{MB}$ and a starting distribution $\pi$ over states at time $t_{start}$ ($\pi$ here is a conditional distribution specified by the belief network $\mathcal{B}$ and the initial states of the other nodes). Unlike the Gibbs sampler of [El-Hay et al., 2008], because our algorithm uses auxiliary variables, our new sample path $\tilde{\mathbf{S}}(t)$ is not completely independent of the previous path, which we denote as $\mathbf{S}(t)$. Recall that $\mathbf{S}(t)$ can be characterized by a sequence of jump times $T$ and states $S$.

Recall that over the time interval, the parents of node $k$ can change states, consequently the rate matrix governing the MJP for node $k$ changes in a piecewise constant manner. Let $A^t$ be the rate matrix at time $t$, and choose some piecewise constant $\Omega^t > \max_s(-A_s^t)$. In the experiments we set $\Omega_t = \max_s(-2A_s^t)$, allowing our subordinating Poisson process to adapt to the dynamics dictated by the current configuration of node $k$'s parents. Once again, we sample an auxiliary set $U$ of virtual jump times, but now from an inhomogeneous Poisson process with rate $\Omega^t + A^t$. Note that the Poisson intensity is still piecewise constant, changing only when either $\mathbf{S}(t)$ changes state (the times in $T$) or when one of the parents changes state (we call this set of times $P$). As before, we will thin out the set $T, U$ by constructing a subordinated Markov chain on the set of times $T \cup U \cup P$. It is important to realize that the MJP for node $k$ will, with probability 1, not jump at the times in $P$, while the transition probabilities at other times are piecewise homogeneous. Thus, at times $t \in T \cup U$, the transition matrix is $B^t = I + \frac{1}{\Omega^t} A^t$, while when $t \in P$ it is simply $B^t = I$.

Order the times in $T \cup U \cup P$ as $t_0 = t_{start} < t_1 < t_2 < \cdots$. Given $(S, T, U)$, if node $k$ had no children, we proceed by resampling the states of the subordinated hidden Markov model using the likelihood function $L_i(s_i)$ in (8). To account for the presence of children $\mathcal{C}$ of node $k$, we have to include the probabilities of the children's sample paths given the complete path $\tilde{\mathbf{S}}(t)$ of node $k$ (as well as those of their other parents). Thus, the likelihood function at step $i$ of the hidden Markov model is:

$$\tilde{L}_i(s_i) = L_i(s_i) \prod_{c \in \mathcal{C}} p(\mathbf{S}^c(t)|\tilde{\mathbf{S}}(t) = s_i, \mathcal{P}^c) \qquad (9)$$

where $t$ ranges over the whole interval $[t_i, t_{i+1})$, over which the sample path of node $k$ is constant and has the value $s_i$, and $\mathcal{P}^c$ denotes the paths of the other parents of $c$. Given its parents' paths, the child $c$'s path over $[t_i, t_{i+1})$ is distributed according to a MJP with piecewise constant rate matrices, since the paths in $\mathcal{P}^c$ could jump during the interval. Thus the probability $p(\mathbf{S}^c(t)|\tilde{\mathbf{S}}(t) = s_i, \mathcal{P}^c)$ in (9) is simply a product of terms, one over each interval of constant rate matrix, and where each term is given by the the probability (5) of a homogeneous MJP (ignoring the initial distribution).

With these likelihood terms, we can again use the forward filtering-backward sampling algorithm to obtain a new state sequence $V$ of node $k$ given the times $W \equiv T \cup U$, giving our new path $\tilde{\mathbf{S}}(t)$ for node $k$. Since $\tilde{\mathbf{S}}(t)$ is obtained via introducing auxiliary variables and performing conditional sampling in the extended space, the MCMC sampler retains the posterior distribution over the sample paths of all nodes as its stationary distribution. Ergodicity is again straightforward to see, so that we have the proposition:

**Proposition 4.** *The auxillary variable Gibbs sampler described above converges to the posterior distribution over the CTBN sample paths.*

## 5 Experiments

In the following, we evaluate a C++ implementation of our algorithm on a number of CTBNs. In all experiments, for a rate-matrix $A$, the parameter $\Omega$ was set to $\max_s(-2A_s)$. The primary question we address is how our sampler (called *Uniformization* in the following) compares to the Gibbs sampler of [El-Hay et al., 2008] for different CTBNs. For this comparison, we used the CTBN-RLE package of [Shelton et al., 2010] (also implemented in C++). In all our experiments, we found this implementation to be significantly slower than our algorithm, especially for large inference problems. To prevent details of the two implementations from clouding the picture (their code also supports parameter and structure learning), we also measured the amount of time the latter spent simply calculating matrix exponentials. This constituted between 10% to 70% of the total running time of the algorithm. In the plots we refer to this as 'El Hay et al. (Matrix Exp.)'. We found that our algorithm took less time than even this, with the speed-up largest for larger problems.

### 5.1 The Lotka-Volterra process

First, we apply our sampler to the Lotka-Volterra process investigated by [Opper and Sanguinetti, 2007]. Commonly referred to as the predator-prey model, this describes the evolution of two interacting populations of 'preys' and 'predators'. The two species form the two nodes of a cyclic CTBN, whose states $x$ and $y$ represent the sizes of the prey and predator populations. The process rates are given by

$$f_{prey}(x+1|x,y) = \alpha x \qquad f_{prey}(x-1|x,y) = \beta xy$$
$$f_{predator}(y+1|x,y) = \delta xy \qquad f_{predator}(y-1|x,y) = \gamma y$$

where the parameters are set as follows: $\alpha = 5 \times 10^{-4}, \beta = 1 \times 10^{-4}, \gamma = 5 \times 10^{-4}, \delta = 1 \times 10^{-4}$. This defines two infinite sets of infinite-dimensional conditional rate matrices. Like [Opper and Sanguinetti, 2007], we limit the maximum number of individuals of each species to 200, leaving us with 400 200-dimensional matrices. Note that these matrices are tridiagonal and very sparse; at any time the size of each population can change by at most one. Consequently, the complexity of our algorithm scales *linearly* with the number of states. A 'true' path of predator-prey population sizes was sampled from this process, and its state at time $t = 0$ was observed noiselessly. Additionally 15 noisy observations were generated, and spaced uniformly at intervals of 100 from $t = 100$ onwards. The noise process is:

$$p(X(t)|\mathbf{S}(t)) \propto \frac{1}{2^{|X(t)-\mathbf{S}(t)|} + 10^{-6}} \qquad (10)$$

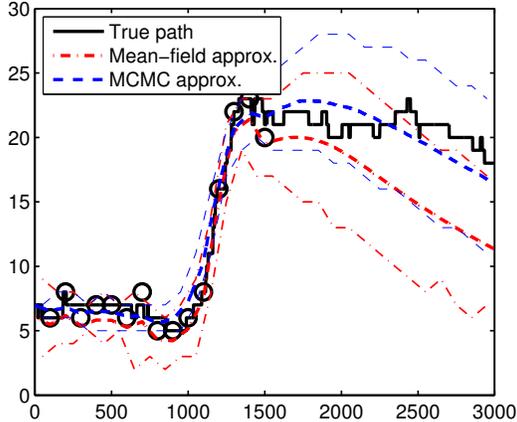

Figure 1: Posterior (mean and 90% confidence intervals) over predator paths (observations (circles) only until $1500$).

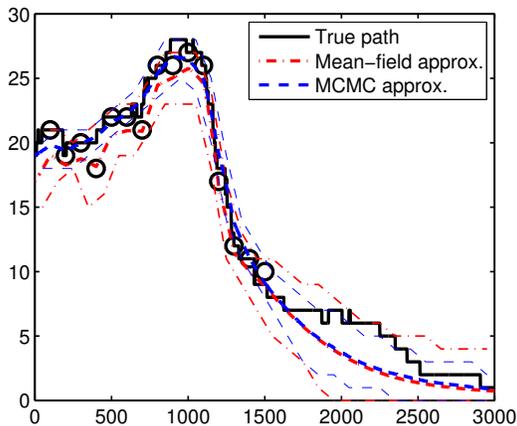

Figure 2: Posterior (mean and 90% confidence intervals) over prey paths (observations (circles) only until $1500$).

Given these observations (as well as the true parameter values), we approximated the posterior distribution over paths by two methods: using 1000 samples from our uniformization-based MCMC sampler (with a burn-in period of 100) and using the mean-field (MF) approxima-tion of [Opper and Sanguinetti, 2007][2]. We could not apply CTBN-RLE to a state-space and time-interval this large. Figures 1 and 2 show the true paths (in black), the observations (as circles) as well as the posterior means and 90% confidence intervals produced by the two algorithms. As can be seen, both algorithms do well over the first half of the interval where data is present. In the second half, the MF algorithm appears to underestimate the predicted size of the predator population; on the other hand, the MCMC posterior reflects the truth better.

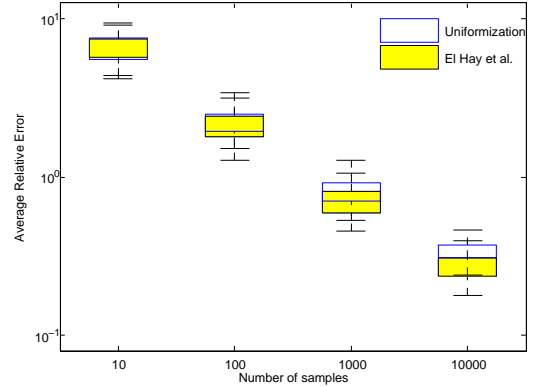

Figure 3: Average relative error vs number of samples for 1000 independent runs; burn-in = 200. Note that in this scenario Uniformization was about 12 times faster, so that for the same computational effort it produces significantly lower errors.

### 5.2 Average relative error vs number samples

In our next experiment, we followed [El-Hay et al., 2008] in studying how *average relative error* varies with the number of samples from the Markov chain. Average relative error is defined by $\sum_j \frac{|\hat{\theta}_j - \theta_j|}{\theta_j}$, and measures the total normalized difference between empirical ($\hat{\theta}_j$) and true ($\theta_j$) averages of sufficient statistics of the posterior. The statistics in question are the time spent by each node in different states as well as the probabilities of transitioning from one state to another. The exact values were calculated by numerical integration when possible, otherwise from a very long run of CTBN-RLE.

As in [El-Hay et al., 2008], we consider a CTBN with the topology of a chain, consisting of 5-nodes, each with 5 states. The states of the nodes was observed at times 0 and 20 and we produced posterior samples of paths over the time interval $[0, 20]$. We calculate the average relative error as a function of the number of samples, with a burn-in of 200 samples. Figure 3 shows the results from running 1000 independent chains for both samplers. Not surprisingly, the sampler of [El-Hay et al., 2008], which does not use auxiliary variables, has slightly lower errors. However the difference in relative errors is minor, and seems negligible when considering the dramatic (sometimes up to two orders of magnitude; see below) speed improvements of our algorithm. For instance, to produce the 10000 samples, the [El-Hay et al., 2008] sampler took about 6 minutes, while our sampler ran in about 30 seconds.

---
[2]We thank Guido Sanguinetti for providing us with his code

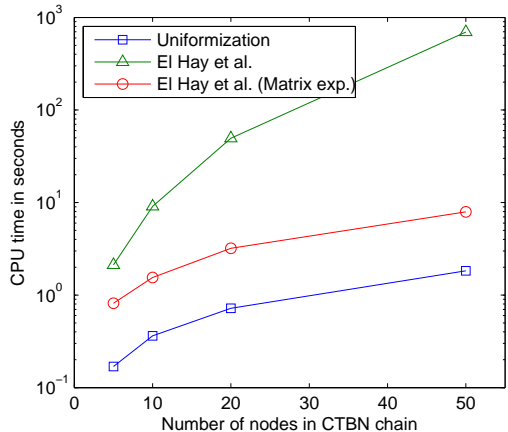

Figure 4: CPU time vs length of CTBN chain.

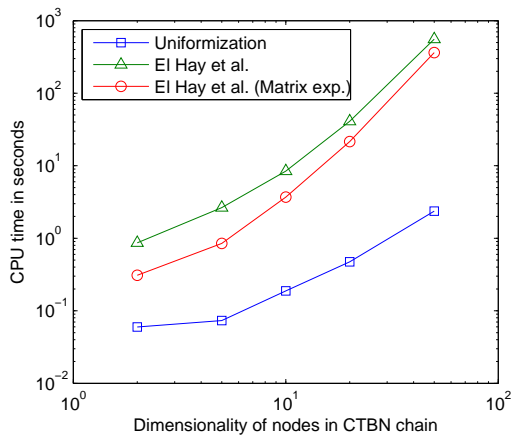

Figure 5: CPU time vs number of states of CTBN nodes.

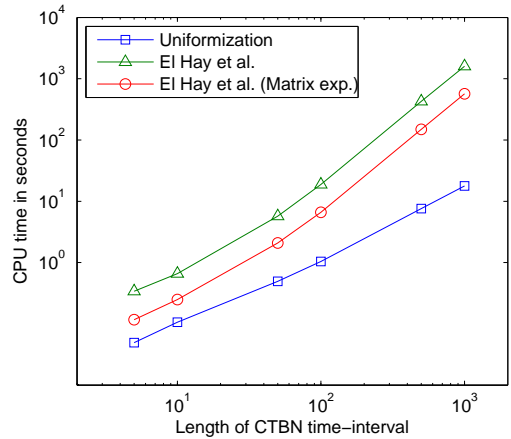

Figure 6: CPU time vs time interval of CTBN paths.

### 5.3 Time requirements

In the three experiments below, we compare the times required by CTBN-RLE and our uniformization-based sampler to produce an effective sample size of a 100 ( calculated using R-CODA [Plummer et al., 2006]). In no case was the actual number of samples produced by our sampler more than three times the number required by CTBN-RLE (suggesting fast mixing). The average relative errors were always comparable for these two sample sizes. All our simulations were run on a laptop with a 2.53 GHz Intel(R) Core(TM) 2 Duo processor and 2GB of memory.

In the first of this set of experiments, we measured the times to produce these samples for the chain-shaped CTBN described above, as the number of nodes in the chain increases. Figure 4 shows the results. The time requirements of both algorithms grow linearly with the number of nodes. However, our uniformization-based sampler always takes less time, and as its rate of growth is smaller; this can become quite large for large chain lengths.

Our next experiment keeps the length of the chain fixed at 5, instead increasing the number of states per node. Once again, uniformization is always faster. From the asymptotic slopes of the two lines, we can verify that its complexity is $O(n^2)$, where $n$ is the number of states, as opposed to $O(n^3)$ for [El-Hay et al., 2008].

Our final experiment measures the time required as the length of the time interval increases. Again, our algorithm is the faster of the two. It is worth pointing out here that algorithm of [El-Hay et al., 2008] has a 'precision' parameter, and that by reducing the desired temporal precision, faster performance can be obtained. However, since our sampler produces *exact* samples (up to numerical precision), we feel our comparison is fair. In the experiments, we left this parameter at its default value.

## 6 Discussion

We proposed a novel Markov chain Monte Carlo sampling method for Markov jump processes and continuous-time Bayesian networks. Our method uses the idea of uniformization, which constructs a Markov jump process by subordinating a Markov chain to a Poisson process. Our auxiliary variable Gibbs sampler is computationally very efficient as it does not require time discretization, matrix exponentiation or diagonalization, and can exploit structure in the rate matrix. In the experiments, we showed significant speed-ups compared to a state-of-the-art sampler for CTBNs, and that our method converges extremely quickly.

Our approach of introducing auxiliary variables to simplify posterior computations in complex stochastic processes bears a relationship to recent MCMC samplers for Gaus-sian processes and Poisson processes [Adams et al., 2009, Murray and Adams, 2010], though the actual techniques used are quite different. We believe these algorithms as well as ours demonstrate how with some thought efficient MCMC algorithms can be developed for important classes of stochastic processes. Another interesting relation to our work is the embedded HMM of [Neal et al., 2004], which uses a hidden Markov model to sample from a discrete time continuous valued Markov process, while ours uses one to sample from a continuous time discrete valued process.

We are currently exploring a number of generalizations of our technique. Firstly, our technique should be applicable to inhomogeneous Markov jump processes where techniques based on matrix exponentiation cannot be applied. Secondly, we can explore Markov modulated Poisson processes which have richer likelihood models than those we considered here. Thirdly, we can explore Markov jump processes with a countably infinite state space, by combining our technique with slice sampling.